\documentclass[]{article}
\usepackage{epsfig}

\newcommand{\be}{\begin{equation}}
\newcommand{\ee}{\end{equation}}
\newcommand{\bea}{\begin{eqnarray}}
\newcommand{\eea}{\end{eqnarray}}
\newcommand{\no}{\nonumber}

\begin{document}

\title{{CP violating neutrino oscillation \\
          and uncertainties in Earth matter density}}

\author{ 
Lian-You Shan$^{a,b}$
\footnote{shanly@ihep.ac.cn}, 
Bing-Lin Young$^c$ , 
~and ~Xinmin Zhang$^a$ \\
$^a$ Institute of High Energy Physics, \\
Chinese Academy of Sciences, P.O.Box 918, Beijing 100039, China   \\
$^b$CCAST (World Laboratory), P.O.Box 8730, Beijing 100080, China  \\
$^c$Department of Physics and Astronomy, Iowa State University \\
Ames, Iowa 50011, U.S.A.  
}
\maketitle 
\begin{abstract}
We propose a statistical formulation to estimate possible errors in long 
baseline neutrino oscillation experiments caused by uncertainties in the 
Earth matter density.  A quantitative investigation of the effect is made 
on the CP asymmetry in future neutrino factory experiments. 

\end{abstract}

Leptonic CP violation (CPV) is one of the main challenges in future 
long baseline (LBL) neutrino oscillation experiments \cite{lbl}, where 
more accurate measurements of the neutrino oscillation parameters 
are anticipated.  However, since the neutrino beam travels a long path 
through the Earth, the MSW matter effect \cite{MSW} can mimic a 
non-vanishing CP phase and makes it non-trivial to extract the intrinsic 
CP phase.  Therefore, a thorough quantitative understanding of the 
matter effect and its possible uncertainties is necessary before an 
accurate account of the CPV effect can be achieved.

The critical quantity in discerning the matter effect is the value of the 
Earth matter density as a function of the baseline.  A number of 
approaches have been suggested on using different Earth density 
profiles of the Earth for the electron number density (DEN).  In 
these works, DEN is taken either as a distance-averaging effective 
constant ~\cite{const}, an adiabatic approximation 
profile~\cite{adiabatic}, mantle-core-mantle layers 
approximation~\cite{mcm3}, multi-step functions~\cite{multis}, 
or the preliminary reference Earth model (PREM)~\cite{PREM}. 
We refer to Ref.~\cite{ohlrev} for a review of some of the 
available Earth density models.  It is clear from these works that 
the analysis of the leptonic CPV will depend on the Earth density 
model adopted to analyze the experimental data.  


From the geophysics point of view, a density 
model is understood to be an approximation of the Earth
and has its inherent uncertainties.
For a brief discussion on the uncertainty of DEN, we refer to Refs. 
\cite{ohpe}
and \cite{mcm3}. Detailed discussions can be
found in geophysics review\cite{invers,bullen,bolt}.
As to PREM, significant deviations from PREM due to local variation 
have been  
documented\cite{cases} and
averaged per spherical 
shell with thickness of 100 Km, its precision is roughly $5\%$.

Instead of examining quantitatively all the schemes of the Earth density to 
find out the preferred density scheme to use, we would like to raise and 
attempt to provide at least a partial answer to the following basic questions: 
What is the effect of uncertainties in the Earth matter density on the
measurement of CPV? 
Can the effect be so severe that it demands a more accurate knowledge 
of the Earth density before a reliable CP phase can be extracted in LBL 
experiments?
  
We propose in this paper to estimate the error due to the uncertainty in the 
Earth matter density as the variance of a CP odd oscillation probability.
As described below the procedure in making the estimate is a weighted 
average over the whole sample space of possible Earth density profiles.  
We first introduce a local variance function $\sigma(x)$ to characterize 
the density uncertainties at each point of the Earth along 
a given baseline.  
With the variance function we sample the earth density profiles, then 
construct an averaging process 
to calculate the deviation of a physical quantity from its mean value.
The probability of the various DEN examples is taken as a logarithmic 
distribution suitable for non-negative quantities, although other statistics 
approach may be used. 

We begin the formulation with the flavor Hamiltonian that governs 
the neutrino propagation in matter.  Omitting terms that leads to only 
a common phase to all flavor states, we have, in the scheme of 
three flavors of neutrinos,
\be
H[ \delta_{cp},N_e(x) ] = \frac{U}{2 E_\nu}
\left( \begin{array}{ccc}
m^2_1 & 0 & 0 \\
0 & m^2_2 & 0 \\
0 & 0 & m^2_3 
\end{array}
\right)
U^\dagger +
\left( \begin{array}{ccc}
\sqrt{2} G_F N_e(x) & 0 & 0 \\
0 & 0 & 0 \\
0 & 0 & 0 
\end{array} \right) ,   
\label{heisenberg}
\ee 
where $E_\nu$ is the energy of the neutrino, $G_F$ the Fermi constant, 
and $m_j$, $j$=1, 2, 3 are the neutrino mass eigenvalues.  U is the 
three-neutrino mixing matrix in the basis where the charged leptons are 
diagonalized~\cite{pmns, H2B} and $\delta_{cp}$ is the CPV phase 
appearing in the mixing matrix.  $N_e(x) $ is the DEN function that 
determines the matter effect.   For the antineutrino $N_e(x) $ is replaced 
by $-N_e(x)$  and U by its complex conjugated which is equivalent to 
replacing $\delta$ by $-\delta$.  The $\nu_\alpha \rightarrow \nu_\beta$ 
oscillation probability and that for their anti-particles can be written 
succinctly to exhibit the DEN dependence as
\bea
P_{\alpha\beta} &\equiv& 
    P_{\nu_\alpha\to\nu_\beta}(L, E, \delta_{cp},N_e(x) )
={ \big{|} \Big({\cal T} exp\big( -i \int_0^L H
[ \delta_{cp}, N_e(x) ] dx \big) \Big)_{\alpha\beta} \big{|} }^2  ,  \no  \\
P_{{\bar\alpha}{\bar\beta} } &\equiv& 
    P_{{\bar\nu}_\alpha\to{\bar\nu}_\beta}(L, E, \delta_{cp},N_e(x) )
={ \big| \Big({\cal T} exp\big( -i \int_0^L H
[ -\delta_{cp}, -N_e(x) ] dx \big) \Big)_{\alpha\beta} \big| }^2 ,
\label{oscip}
\eea
where ${\cal T}$ denotes a propagation path-ordering product 
\cite{lecturenotes}.  For numerical implementation we follow the
usual approach of numerically integrating the Schr\"{o}dinger
equation.

Although EDN is a critical factor in the analysis of the long baseline 
oscillation data, 
what is usually available is an averaged Earth density function 
${\hat N}_e(x) $ such as the widely used PREM model.  
 In the following we 
study the effect of variations from the average value.  Let us define 
the average density ${\hat N}_e(x)$ and its uncertainties given
by the variance function $\sigma(x)$,
\be
{\hat N}_e(x) =<N_e(x)> = \int [ {\cal D}  N_e  ] N_e(x) F[ N_e(x) ] ,
\no \\
< \sigma (x) > = \sqrt{ < N^2_e(x) > - { <N_e(x)> }^2  } ,
\ee
where $ F[ N_e(x) ] [{\cal D} N_e] $ is the probability of obtaining  
the DEN $N_e(x)$ in the neighborhood $x$, which will be defined in more 
detail later. 
The oscillation probability can be defined as an average over all the 
possible DEN profiles around $ {\hat N}_e(x) $.  The appropriate
framework for such a statistical expectation is the functional integral
in which the Earth density profiles span a functional space
that contains all possible variations of the earth densities allowed 
within the given variance.    We write,
\be
< P_{\alpha\beta} > =
\int [ {\cal D}  N_e  ] P_{\alpha\beta}
 F[ N_e(x) ] . 
\label{PP}
\ee

 In geophysics,  a uniform 
random sampling in a broad space of the Earth matter density are used 
to generate the matter density models \cite{monte,blnk}.  The 
density function generated in this way is further constrained by testing 
against 
two important sets of observational data. One is the mass and moment 
of inertia of the Earth, and the other the normal modes of the free 
oscillation of the Earth.
In general the samples remained approach a Gaussian-like statistics
rather than a uniform one, which together with the fact that the Earth 
matter
density is always positive suggest to us to recast it into a 
logarithmic
normal distribution for the probability density\cite{logau},
\bea
& &F[ N_e(x) , x ] = \frac{1}{ N_e (x) \sqrt{2 \pi} s( x ) }
exp\Big( -  \ln^2 \big( N_e(x) / N_0(x) \big)
 / \big( 2 s^2 ( x ) \big) \Big) \no  \\
 & & s(x) = \sqrt{ \ln \big( 1 + r^2(x)  \big) },~~
 N_0(x) =  {\hat N}_e(x) exp \big( -s^2(x)/2 \big)
 \label{eqlogaus}
 \eea
where $r(x) = \sigma (x) /{\hat N}_e(x)$ parameterizes the uncertainty in     
DEN in terms of the ratio of local variance and the local mean value of
the Earth's density.

The logarithmic distribution is not a symmetric distribution for
arbitrary $\sigma$.  However, it is close to the Gaussian distribution 
for $\sigma$ small in comparison with $\hat{N}_e$. 
In Fig. 1 we plot the logarithmic and Gaussian 
distributions for $r=5\%, 50\%$ respectively with $\hat{N}_e(x)$ given 
by PREM.  We see that the logarithmic distributions 
$N_e(x)$ is always positive and 
the
difference between the Gaussian and the logarithmic distributions
is very small in the case of $r=5\%$ .
In Fig.2, we plot the density profile, from which we see that the PREM 
denoted by the thick solid line is a average of the geophysical density 
samples shown by the oscillating thin lines.

With the inclusion of uncertainties in DEN, we can now estimate the 
uncertainty in the neutrino oscillation probability by computing the 
variance of the oscillation probability, 
\be
\delta P_{\alpha\beta}
 \equiv \sqrt{ < { ( P_{\alpha\beta} - < P_{\alpha\beta} > ) }^2 > } =
\sqrt{ 
\int { ( P_{\alpha\beta} - <P_{\alpha\beta}>) }^2 F[ N_e(x) ] 
[ {\cal D}  N_e ] }.
\label{Pvar}
\ee 
The variance $\delta P_{\alpha\beta} $ leads to an uncertainty in the number 
of observed charged leptons.   Since the charged lepton events are usually 
divided into energy bins, we define the variance of event number in a bin as, 
\be
\delta N_\beta ( E, L ) = \phi_{\nu_\alpha}(E,L) 
\delta P_{\alpha\beta} 
\sigma_\beta ( E ) \Delta E, 
\label{VarN}
\ee
where $\phi_{\nu_\alpha}(E,L)$ is the neutrino beam flux spectrum 
of flavor $\alpha$, $\sigma_\beta ( E )$ is the charged current cross 
section of neutrino of flavor $\beta$, and $\Delta{E}$ is the bin size.  

To measure the CPV effect, the difference between event rates of 
opposite charged leptons is usually considered, 
\bea
N_{cp} ( E,L,\delta_{cp} ) & \equiv & N_\beta - N_{\bar\beta} =
[ \phi_{\nu_\alpha}(E,L) P_{\alpha\beta} \sigma_\beta ( E ) -  
\phi_{{\bar\nu}_\alpha}(E,L) P_{{\bar\alpha}{\bar\beta}} 
\sigma_{\bar\beta} ( E )  ] \Delta E  \no   \\
&=&\phi_{\nu_\alpha}(E,L)  
D^{CP}_{\alpha\beta}(\delta_{cp} ) 
\sigma_\beta ( E ) \Delta E ,
\label{CPN}
\eea
where the CP-odd difference~\cite{diffe,const,adiabatic} is given by,  
\be
D^{CP}_{\alpha\beta}(\delta_{cp}) \equiv 
P_{\nu_\alpha\to\nu_\beta}(L, E, \delta_{cp} ) - 
P_{{\bar\nu}_\alpha\to{\bar\nu}_\beta}(L, E, \delta_{cp} )
\label{cpd}.
\ee 
In our numerical calculations, we have assumed that the neutrino and 
anti-neutrino beams have the 
same flux spectrum and the mass of the detector for the anti-neutrino 
is twice of that of the neutrino so as to compensate the difference in 
the neutrino and anti-neutrino charged-current cross sections.  

Now the error from the uncertainty of DEN is estimated as the standard 
deviation, 
\bea
 \delta D^{CP}_{\alpha\beta} & \equiv & \sqrt{ < { [
   P_{\alpha\beta} -  P_{{\bar\alpha}{\bar\beta} } 
- < P_{\alpha\beta} - P_{{\bar\alpha}{\bar\beta} } > 
   ]  }^2  > }    \no  \\
& = & \sqrt{ 
{ [ \delta P_{\alpha\beta} ] }^2 
+ { [ \delta P_{{\bar\alpha}{\bar\beta} } ] }^2 .
 }
\label{VarCPN}
\eea
Unless this uncertainty is under control it will be difficult to extract 
the CP phase. It may even be difficult to establish the CPV effect if 
the fluctuation caused by the uncertainty of the matter density is not 
much smaller than the typical CP asymmetry given by,
\be
\Delta D^{CP}_{\alpha\beta} \equiv 
D_{\alpha\beta}(\delta_{cp}) -
D_{\alpha\beta}(\delta_{cp} = 0 ) .
\label{CPA}
\ee
Note that we can also work with the normalized conventional 
CP asymmetry,
\be
A^{cp} = \frac{ D^{CP}_{\alpha\beta} }
{P_{\alpha\beta}+ 
   P_{{\bar \alpha}{\bar \beta} } }
,~~~ \delta A^{cp} = 
\sqrt{ 
\int { ( A^{cp} - <A^{cp}>) }^2 F[ N_e(x) ] [ {\cal D}  N_e ] },
\ee
which should provide the same information.  

We evaluate Eqs. (4) and (6) numerically using a method similar to that 
of the lattice gauge theory.
The neutrino path is discretized into $I$ one-dimensional cells where 
$I$ is a sufficiently large integer.  In each of the $ i$-th cell, the DEN 
function $ N_e( x_i )$ has an independent logarithm normal distribution 
with its local variance $\sigma(x_i)$.  Then the estimators of the mean 
and the deviation can be recasted respectively into the following forms,
\bea
< D^{CP}_{\alpha\beta} > &=& \lim_{I\to\infty} \int 
D^{CP}_{\alpha\beta}[ \{\Delta m^2 , \theta; \delta_{cp} \};
\{ N_e( x_1 ), ... N_e( x_i ), ... N_e( x_I) \} ] 
\prod_{i=1}^I F[ N_e(x_i),x_i ] 
\frac{ d N_e( x_i ) }{ N_e(x_i) \sqrt{ 2\pi }s(x_i) }  \no \\ 
& = & \lim_{K\to\infty} K^{-1} \sum_{k=1}^K 
{ \tilde D_k } [ \{\Delta m^2 , \theta , \delta_{cp}  \}; 
{ \{ N_e \} }_k ]   
\label{dd}  \\
\delta D^{CP}_{\alpha\beta} &=& { \{ \lim_{K\to\infty} {(K-1) }^{-1} 
\sum_{k=1}^K { [{ \tilde D_k } - < D^{CP}_{\alpha\beta} > ]}^2 \} }^{1/2}
\label{lattice}
\eea
where
we have replaced the functional integration over the DEN by a sum over
K arrays, $\{N_e\}_k$, $k=1,2,...K$, 
and $ \tilde D_k $ is the value of the difference $D^{CP}_{\alpha\beta}$  
over the $ k$-th density function ${ \{ N_e \} }_{k}$.  The array, 
${ \{ N_e \} }_{k}$ which consists of density function $N_e(x_1), ...
N_e(x_i)...N_e(x_I)$
is generated from PREM weighted with a 
Gausian-like logarithm deviation. Of course, other specific 
Earth density models can be used. We have checked numerically that Eqs. 
(\ref{dd}) and (\ref{lattice}) are convergent and stable.
  
Geophysically, to obtain the density profile, the earth is discretized into 
hexahedrons
( elementary volume in spheroidal coordinates ), then 
Earth density is defined on the nodes and solved as an inverse problem.
Generally there exists a length scale of the 
hexahedrons, which typically is order of 100 Km\cite{invers,logau,uhrh}.
So we take I to be the order of 
$L/ 100$Km in our calculation.  Furthermore since the series in Eqs. (13) 
and (14) converges very rapidly due to the Gaussian-like distribution of 
Eq. (5), for the convenience of the calculation we identify $K$ with the 
number of beam neutrinos in the individual bins, i.e., 
\be
K=K(E,L) = \phi_{\nu_\alpha}(E,L) \sigma_\alpha (E)\Delta{E}.
\label{samples}
\ee


Having constructed all the needed ingredients, we can now investigate 
the error in CPV due the uncertainty of matter density. In the numerical 
calculation we take the baseline 2900 km for illustration. This baseline 
has been widely used for the study of CP violation at neutrino factories. 
We take a 20 GeV high performance neutrino factory, which delivers 
${10}^{21}$ working muons per year to a 50 kiloton detector.
  Since the path 
of the 2900 km baseline can go as deep as 160 km into the Earth and 
this will reach part of the low velocity zone, the uncertainty of the 
PREM can be sizable. In the numerical calculation, we take $r(x) = 5\%$ 
and the lattice size to be 200 km.  

We adopt the large mixing angle (LMA) scenario \cite{sno} for the solar 
neutrino puzzle and take the following typical set of mixing parameters, 
\bea
& &\Delta m^2_{sol}  = 6.0 \times {10}^{-5} {\rm eV^2}, ~~~~
\Delta m^2_{atm} = 3.55 \times {10}^{-3} {\rm eV^2},   \no  \\
& &tan^2 \theta_{12} = 0.3,~~~~  
\hspace{11.5ex} sin^2 2 \theta_{23}= 0.99 ~~~.
\eea
We also take the typical value $sin^2 \theta_{13} = 0.08 $.
 
Now we present the numerical results.   First, we study the extend to 
which the matter effect will mimic the CPV effect.  In Fig.3, we plot 
$N_{cp}$ as a function of the neutrino energy. The dashed line is 
given by $\delta_{cp}=0$ with the error bars for an $5\%$ uncertainty 
in PREM. 
So at this baseline the uncertainty caused by that of the matter density 
seems to be partially controllable.  To see this more clearly and to 
estimate the range of the uncertainty, we show in Fig. 4 the following 
cases:  
dashed curve for $\delta_{cp} = 90^\circ$ with 5\% uncertainty in PREM, 
dotted and solid curves respectively for $\delta_{cp} = 54^\circ$ and
$\delta_{cp} = 0$ without uncertainty in PREM.  It is clear that 
$\delta_{cp} = 90^\circ$ can be easily distinguished from $\delta_{cp} =
0$.
However, there is an large error in extracting the CP phase that is about 
36$^\circ$.   

The uncertainty increases with the baseline because of the 
accumulation of the matter effect.  In Fig. 5 we plot $N_{cp}$ vs the 
neutrino energy for a 12000 Km baseline.  We consider again a neutrino 
factory which delivers $10^{21}$ working muons per year but at 
50 GeV to increase the statistics. One can see that the uncertainty is 
large and it can no longer to distinguish  $\delta_{cp}=90^\circ$ from 
$\delta_{cp}=0^\circ$.  So in order to have sensitivities for the CP 
measurement at this distance, the accuracy in DEN has to be 
much better than $5 \%$.

We note that the effect of uncertainties in the Earth density have also
been examined in Ref. \cite{pinney,mena}.  However the uncertainty 
has been fixed at the maximum value, $\it {i.e.}$, 
$N_e(x) = ( 1 \pm r' ) {\hat N_e(x)}$ where $\hat{N}_e(x)$ is also 
given by PREM.   As a comparison, we show in Fig. 5, together with 
our results, the constant uncertainties with $r'=5\%$ (dot-dashed and 
dashed lines).  
In our language this fixed-value distribution function is represented by 
a delta function.  So we think it overestimates the effects of the 
uncertainty of the Earth density.

To conclude, we have considered in some details the issue of extract
the information of the CPV effect in the presence of uncertainties in 
the Earth matter density in high precision measurements anticipated in 
future LBL neutrino experiments.  We have developed a path integral 
formulation to estimate the fluctuation around the statistical mean.  
We have also presented a numerical implement of the formulation 
and applied it to assess the effectiveness in the determination of the 
CP phase.  

We thank Yi-Fang Wang for discussions.
We also thank Fu-Tian Liu in the National Geological
and Geophysics Institute for geophysics discussion.   
The work is supported in part by the NSF of China
under Grant No 19925523
and also supported by the Ministry of Science and Technology of China
under Grant No NKBRSF G19990754.

\newpage

\begin{figure}[t]
\vspace{1.0cm}
\begin{center}
\epsfig{file=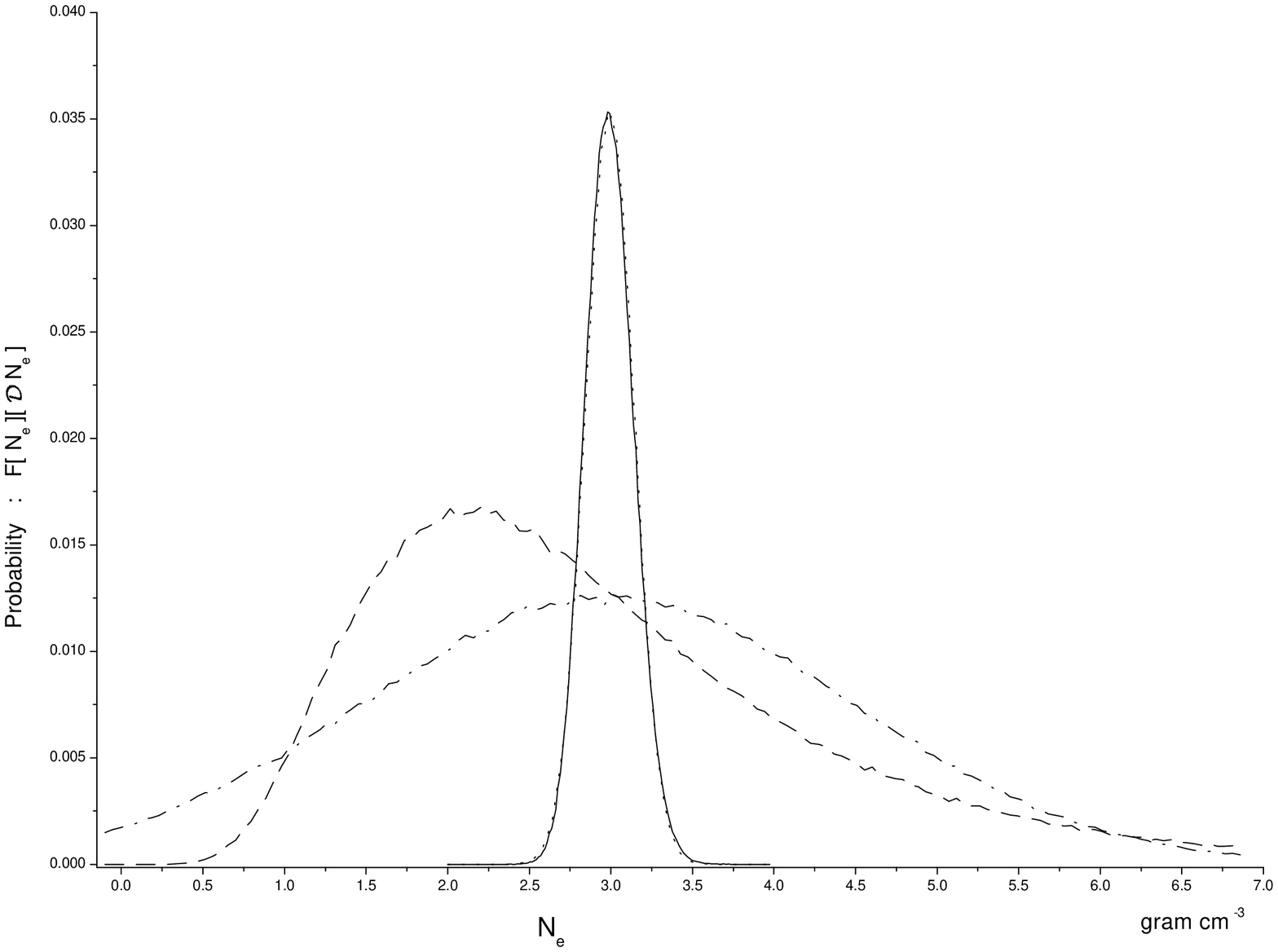,width=18cm}
\caption{ 
Plot of probability distribution $F[ N_e(x)][{\cal D} N_e(x) ]$ 
vs density $N_e(x)$ at depth, {\it e.g.}, 
$ R =6200 $ Km from the earth surface. 
The mean value is taken from PREM which is
$3 g/{cm}^3 $. The solid and dotted lines are the logarithmic and 
the Gaussian distribution with $r=5\%$, which are almostly 
coincided. The dash and dot-dashed lines are the logarithmic and Gaussian 
distribution for $r=50\%$ respectively, which show that Gaussian 
distribution leads to negative density. } \label{flogaus} \end{center}
\end{figure}

\begin{figure}[t]
\vspace{1.0cm}
\begin{center}
\epsfig{file=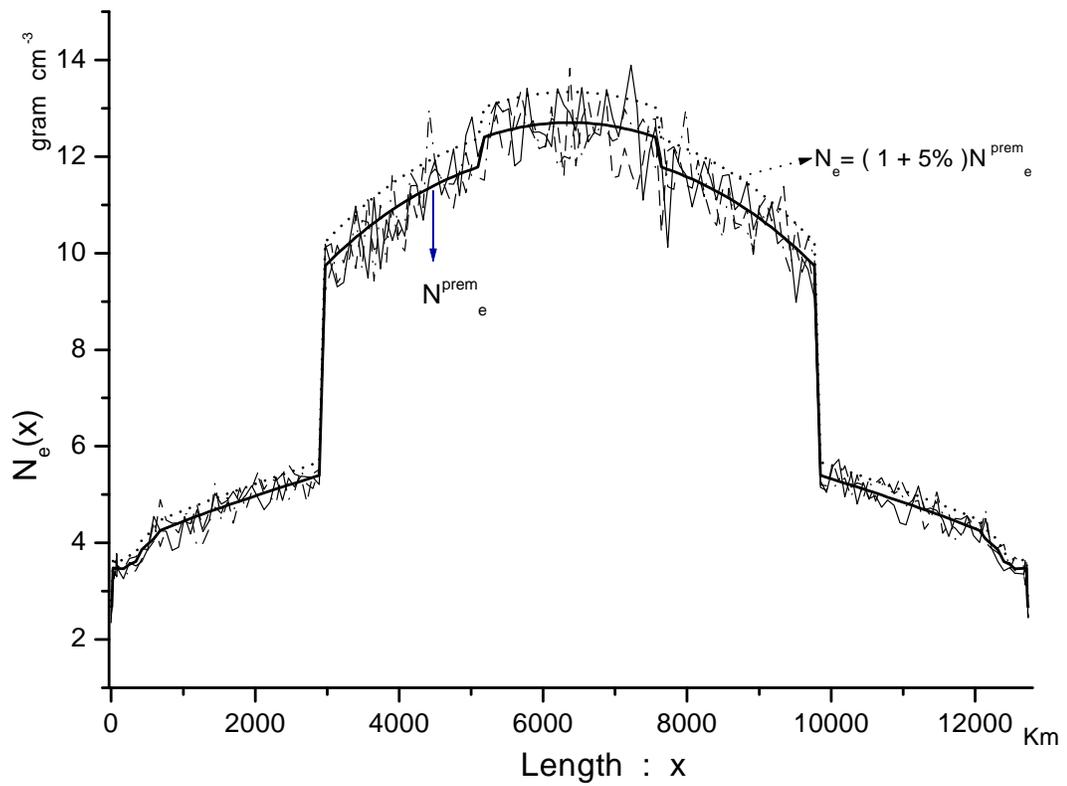 , width=18cm }
\caption{
Plot of earth matter density
vs path length. The thick solid line is the prediction of PREM 
$\hat{N_e(x)}$. The dashed line above PREM is for
$ N_e(x) = ( 1 + 5\% ) {\hat N}_e(x) $. 
The oscillating lines
are sample profiles generated according to 
Eq.(\ref{eqlogaus}) with $r(x)= 5\% $ . }\label{premave}
\end{center}
\end{figure}

\begin{figure}[t]
\vspace{1.0cm}
\begin{center}
\epsfig{file=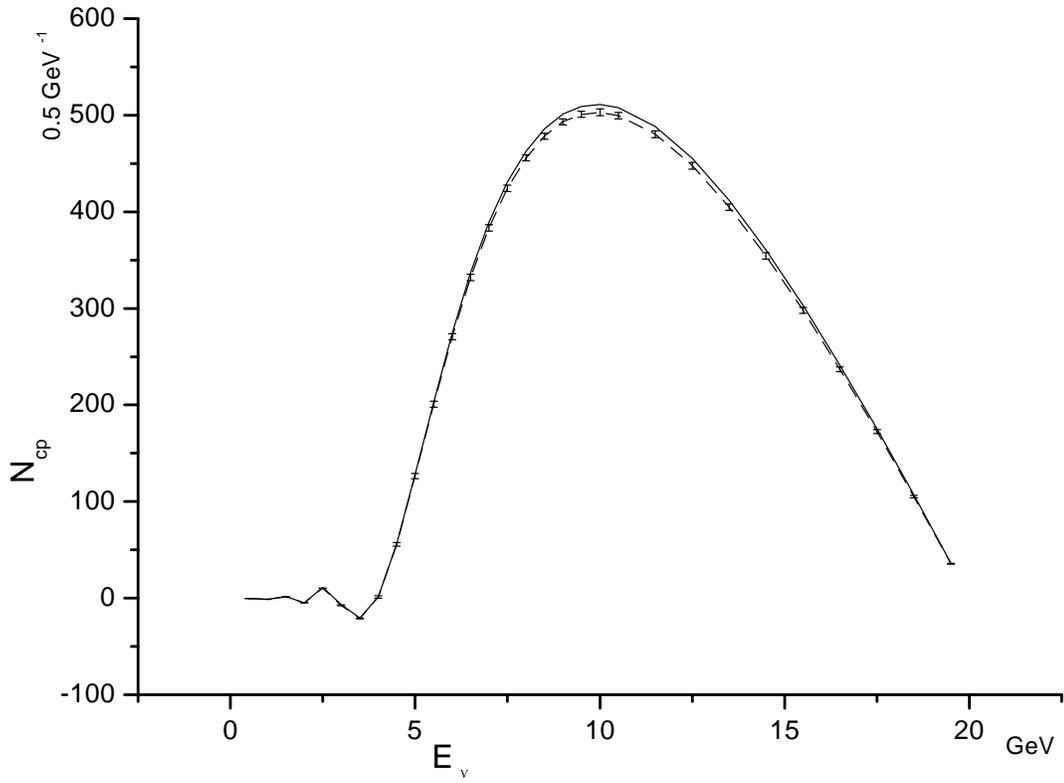,width=18cm}
\caption{ Plot of CP-Odd event number difference $N_{cp}$ 
vs neutrino energy $E_{\nu}$ for baseline L=2900Km. 
The dashed line is for  $\delta_{cp} = 0^\circ $ and the error bars
represents 
the variance caused by the uncertainty of matter density with 
 $r=5\%$.
The solid line is the prediction of
$\delta_{cp} = 7.5^\circ $ in PREM, ${\it i.e.} ~r(x)=0 $ .
}
\label{d29cp5}
\end{center}
\end{figure}

\begin{figure}[b]
\vspace{1.0cm}
\begin{center}
\epsfig{file=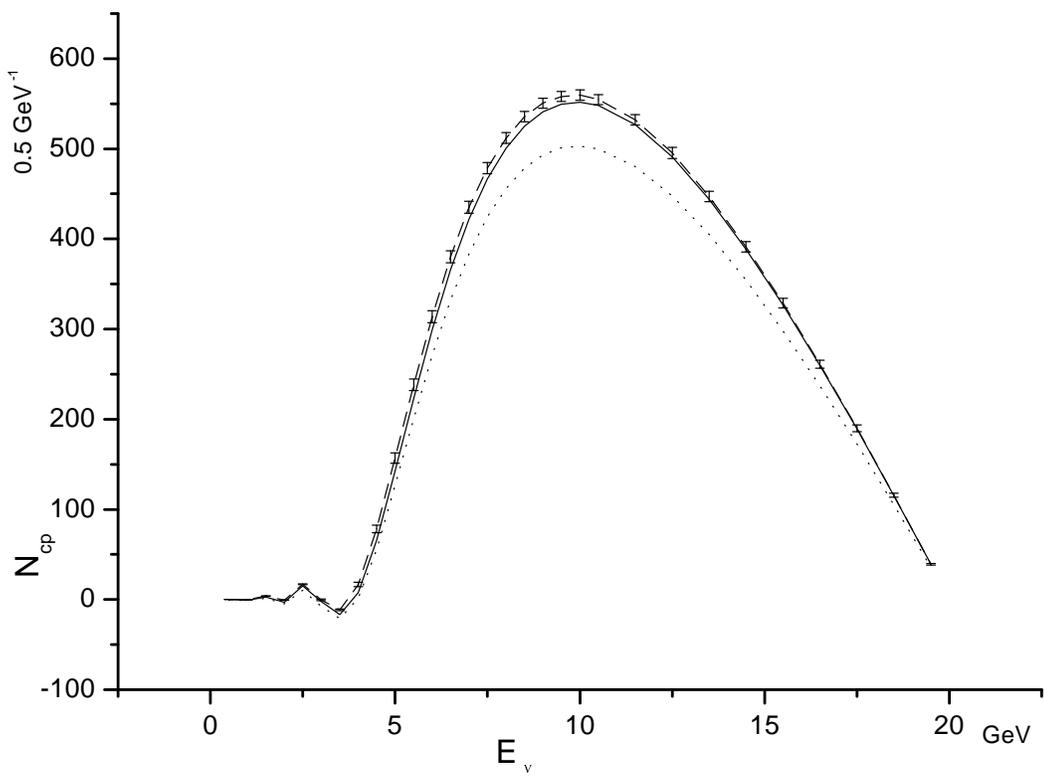,width=18cm}
\caption{ The same plot as Fig.3, but parameters chosen differently.
The dashed line is for
$\delta_{cp}=90^0$ with error bars representing the uncertainties of 
$r=5\%$.
The solid and dotted line are for 
$\delta_{cp}=54^0$ and $ 0^0 $ respectively with $r=0$.
}
\label{d29maxcp}
\end{center}
\end{figure}

\begin{figure}[b]
\vspace{1.0cm}
\begin{center}
\epsfig{file=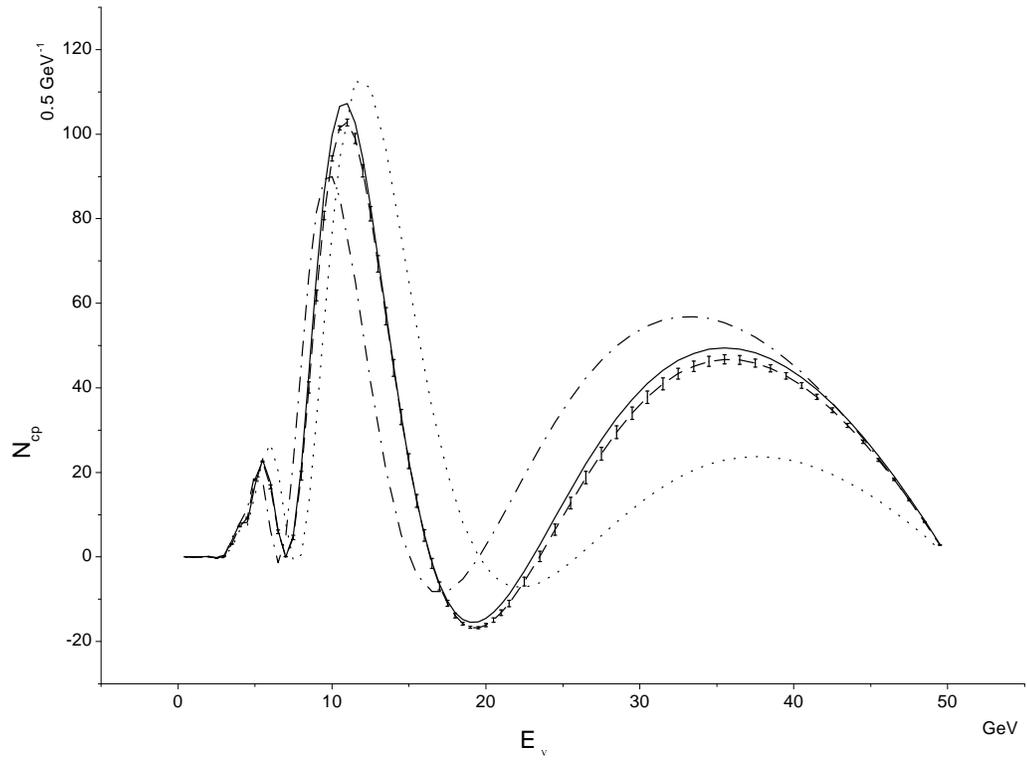,width=18cm}
\caption{ The plot of CP-Odd event number difference $N_{cp}$ 
 vs the neutrino energy $E_{\nu}$ for baseline $L
=12000 $ Km. The dashed line is for $\delta_{cp}=0^0$ with error bars
representing $r=5\%$. The solid line is for $\delta_{cp}=90^0$ with $r=0$.
The dash-dotted and dotted lines are given by $N_e(x) = ( 1 \pm 0.05 ) {\hat 
N_e(x)}$ respectively, where $\hat N_e(x)$ is given by the PREM.  
} \label{dLp5}
\end{center}
\end{figure}

\end{document}